\documentclass[%
preprint,
amsmath,amssymb,
aps,
]{revtex4-2}

\usepackage{amsmath}
\usepackage{amssymb}
\usepackage{graphicx, bm}
\usepackage{dcolumn}
\usepackage{csquotes}
\usepackage{float,soul}
\usepackage{float}
\usepackage{placeins}
\usepackage[none]{hyphenat}
\usepackage{wrapfig}
\usepackage{dblfloatfix}
\usepackage{hyperref,xcolor}
\hypersetup{hidelinks}
\hypersetup{colorlinks=true,linkcolor=black,urlcolor=black}
\usepackage{tabularx}
\usepackage{adjustbox,booktabs,multirow,xfrac}
\begin{document}
	\title{A Benchmark Study of the Relativistic Plane Wave Impulse Approximation for Polarized ($\vec{p},2\vec{p}\,$) Reactions on the 3s$_{1/2}$ state in $^{208}$Pb at 392 MeV}
\textsl{}
\author{T. Mello}
\email{MT17000947@biust.ac.bw}
\author{G. C. Hillhouse}
\author{J. P. W. Diener}
\affiliation{$^1$Department of Physics and Astronomy, Botswana International University of Science and Technology (BIUST), Private Bag 16, Palapye, Botswana}
	\begin{abstract}
	The Relativistic Plane Wave Impulse Approximation (RPWIA) is benchmarked against the established Relativistic Distorted Wave Impulse Approximation (RDWIA) for exclusive proton-induced proton knockout reactions ($\vec{p},2\vec{p}\,$) from the 3s$_{1/2}$ state in $^{208}$Pb at 392 MeV. While the RDWIA provides the proper description of the reaction, the RPWIA—which neglects nuclear distortion and absorption effects—fails dramatically for absolute cross-sections, requiring a scaling factor of $\sfrac{1}{26}$ to match the data. However, within $\pm10$ MeV of the recoilless condition, the RPWIA quantitatively describes the analyzing power ($A_y$) data and yields similar results as the RDWIA predictions for all polarization transfer observables (D$_{i'j}$). These polarization transfer observables are insensitive to the choice of the Relativistic Mean Field model used. The study establishes clear boundaries for the RPWIA's utility: computationally simple predictions for polarization observables near zero recoil, but essential reliance on the RDWIA for spectroscopic factor extraction.
	\end{abstract}
	
	\maketitle
	
\section{Introduction}
The study of nuclear reactions with polarized beams, particularly exclusive ($\vec{p},2\vec{p}\,$) reactions, provides insights into nuclear structure and reaction mechanisms. These \newline experiments have been instrumental in determining nucleon momentum distributions, spectroscopic factors, and the medium-modification of the nucleon-nucleon ($NN$) interaction \cite{PhysRevC.27.1060,PhysRevC.67.064604,Noro2020ExperimentalSO,PhysRevC.66.034602}. The most sophisticated theoretical descriptions of these reactions are provided by distorted wave models which incorporate the effects of the nuclear medium through optical potentials that describe the distortion of the incident and outgoing proton waves. These potentials can be derived phenomenologically by fitting elastic scattering data \cite{PhysRevC.47.297,PhysRevC.68.034608} or microscopically through folding models \cite{Horowitz1991,Bartover}.\\
    \\
The Non-Relativistic Distorted Wave Impulse Approximation (NRDWIA) has achieved broad success in spectroscopic analyses of knockout reactions for a wider range of nuclei and single-particle states \cite{Noro2020ExperimentalSO,WAKASA201732}. For lighter nuclei such as $^{12}$C, $^{16}$O and $^{40}$Ca, the NRDWIA model provides spectroscopic factors which are consistent with those extracted from the electron-induced knockout ($e,e'p$) measurements \cite{Noro2020ExperimentalSO}. However, for heavier nuclei such as $^{208}$Pb, NRDWIA was found to overestimate the spectroscopic factors as compared to the ($e,e'p$) which suggests that additional effects may become important in heavier nuclei \cite{Noro2020ExperimentalSO}. The Relativistic Distorted Wave Impulse Approximation (RDWIA) is a fully relativistic framework that treats the scattering and the boundstate wave functions within the Dirac theory, and its applications have been focused largely on medium modifications of the $NN$ interaction and spin-dependent observables where the relativistic effects are expected to be significant \cite{PhysRevC.68.034608,PhysRevC.74.064608,PhysRevC.67.064604}.\\
   \\
	A known challenge in distorted-wave approaches, whether non-relativistic or relativistic, is their sensitivity to the choice of optical potentials, which can introduce ambiguity in the extraction of absolute spectroscopic factors \cite{Noro2020ExperimentalSO,COWLEY1995300}. This has motivated investigations into simpler models to identify observables and kinematics that are not sensitive to optical potentials. We therefore consider to investigate whether a \enquote{simpler}, Relativistic Plane Wave Impulse Approximation (RPWIA) model which treats all proton wave functions as Dirac plane waves that neglect the distortion and absorption effects \cite{Mello,PhysRevC.69.024618}—can describe analyzing powers for ($\vec{p},2\vec{p}$\,) reactions under specific kinematic regimes such as near the recoilless condition. \\
	\\
	This paper presents a benchmark of the RPWIA model against the established RDWIA standard.~We do not propose RPWIA as a replacement for sophisticated distorted-wave models. Instead, we seek to quantitatively map its performance across a complete set of observables—the unpolarized triple differential cross section ($\sigma$), the analyzing power (A$_{y}$), and the full set of polarization transfer observables (D$_{i'j}$)—for proton knockout from the 3s$_{1/2}$ state in $^{208}$Pb. This paper aims to:
	\begin{enumerate}
		\item Quantify the failure of the RPWIA to describe the absolute cross-section.
		\item Identify the specific kinematic region where its predictions for A$_{y}$ and D$_{i'j}$ coincide with the RDWIA and experimental data.
		\item Investigate the sensitivity of these findings to the underlying nuclear structure model.\newpage
        \item Compare the spectroscopic factor for the ($\vec{p},2\vec{p}$\,) reaction extracted from the RDWIA for the 3s$_{1/2}$ state in $^{208}$Pb with NRDWIA analyses and ($e,e'p$) measurements.
	\end{enumerate}
This analysis provides a clear reference for the utility and limitations of the RPWIA model, which may be of value for exploratory calculations, and for informing future ($\vec{p},2\vec{p}$\,) experiments on stable and unstable nuclei, particularly those focusing on polarization transfer observables.\\
    \\
This manuscript is organized as follows. Section \ref{Theoretical-framework} outlines the theoretical framework, including the RPWIA formalism (Sec.~\ref{Sec:RPWIA}), the relativistic boundstate wave functions (Sec.~\ref{Sec:Relativistic boundstate wave functions}), and the Nucleon-Nucleon ($NN$) interaction with the IA1 parameterization (Sec.~\ref{Sec:Nucleon-nucleon}). Section~\ref{Sec:Results and Discussion} presents the results and discussion for the triple differential cross section, analyzing power and polarization transfer observables for the 3s$_{1/2}$ state of $^{208}$Pb. Our conclusions with a discussion of the utility and limitations of the RPWIA framework are presented in Section~\ref{sec:conclusion}.
	\section{Theoretical Framework}\label{Theoretical-framework}
	\subsection{Relativistic Plane Wave Impulse Approximation}\label{Sec:RPWIA}
\begin{figure}[htbp]
    \centering
    \includegraphics[width=0.98\textwidth]{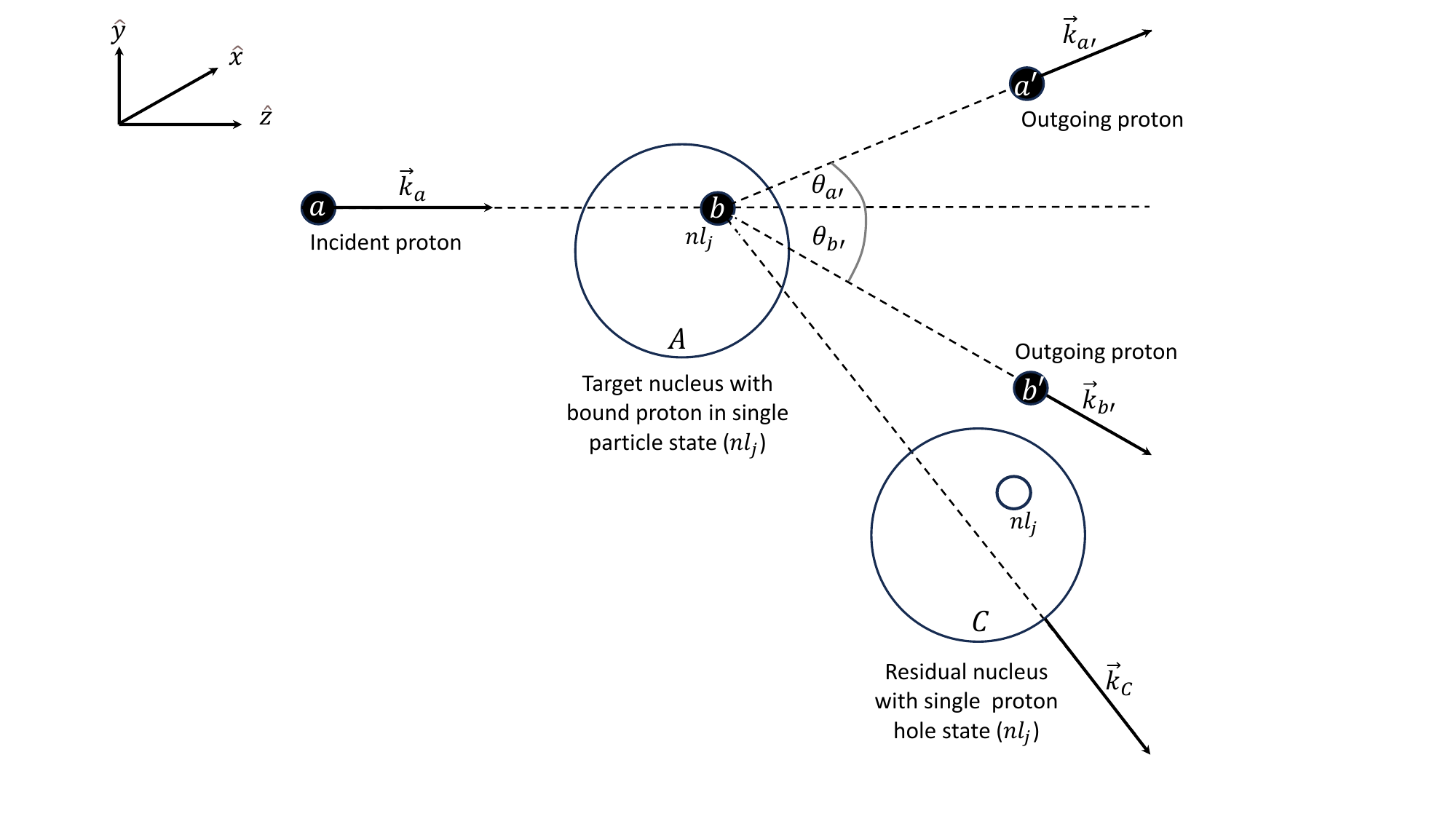}
    \caption{Schematic diagram of the exclusive $(\vec{p},2\vec{p}\,)$ knockout reaction. }
    \label{fig:p2pgeometry}
\end{figure}

	The exclusive polarized proton-induced proton-knockout reaction ($\vec{p}, 2\vec{p}\,$), illustrated by Fig.~\ref{fig:p2pgeometry}, involves a polarized projectile proton ($a$) with momentum $\vec{k}_{a}$, knocking out a bound proton ($b$) from a single-particle state ($n {l}_{j}$) of a stationary target nucleus ($A=C+b$) leaving behind the residual nucleus $C$, such that two outgoing protons $a'$ and $b'$ with momenta $\vec{k}_{a'}$ and $\vec{k}_{b'}$ respectively, are detected at coplanar scattering angles $\theta_{a'}$ and $\theta_{b'}$ in the exit channels \cite{Hillhouse:2006dj,PhysRevC.69.024618}. The principal quantum number, the orbital angular momentum and the total angular momentum are given as $n$, ${l}$ and $j = l\pm\frac{1}{2}$, respectively.
 The transition amplitude for this particular reaction is given by:
	\begin{multline}
		T_{ljm_{j}} (s_{a},s_{a^{\prime}}, s_{b'}) = \\ \int d^{3}\vec{x}\, \big[
		\bar{\psi}^{(-)}(\vec{x},\vec{k}_{a^{\prime}}, s_{a^{\prime}})
		\otimes 
		\bar{\psi}^{(-)}(\vec{x},\vec{k}_{b'},s_{b'}) 
		\big] 
		\times 
		\hat{F} \big[
		\psi^{(+)}(\vec{x},\vec{k}_{a},s_{a}) 
		\otimes 
		\phi_{ljm_{j}}(\vec{x}\,)
		\big],
		\label{Transition-Amplitude}
	\end{multline}
	where the wave functions are treated as relativistic Dirac plane waves:
	\begin{equation}
		\begin{aligned}
			\psi^{(+)}(\vec{x},\vec{k}_{a}, s_{a}) &= e^{i\vec{k}_{a}\cdot\vec{x}}U(\vec{k}_{a},s_{a}),\\
			\psi^{(-)}(\vec{x},\vec{k}_{a'}, s_{a'}) &= e^{-i\vec{k}_{a^{\prime}}\cdot\vec{x}}U(\vec{k}_{a^{\prime}},s_{a^{\prime}}),\\
			\psi^{(-)}(\vec{x},\vec{k}_{b'}, s_{b'}) &= e^{-i\vec{k}_{b'}\cdot\vec{x}}U(\vec{k}_{b'},s_{b'}).
		\end{aligned}\label{EQ:Free Dirac Plane Wave}
	\end{equation}
	The four-component Dirac spinor is given as:
	\begin{equation}
		U(\vec{k}_{i},s_{i}) = \sqrt{\frac{E_{i} + m_{i}}{2m_{i}}}
		\left( \begin{array}{c}1 \\
			\frac{\vec{\sigma} \cdot \vec{k}_{i}}{E_{i} + m_{i}}
		\end{array}   \right)\chi_{s_{i}}.
	\end{equation}
with the Pauli spin matrices denoted by $\vec{\sigma}$, the energy and the rest mass of the nucleon by $E_{i}$ and $m_{i}$ respectively. The two-component Pauli spinors with projection $s$ are represented by $\chi_{s_{i}}$. The wave function of the incident proton is represented as $\psi^{(+)}(\vec{x},\vec{k}_{a},s_{a})$, while the wave functions of the two outgoing protons are denoted as $\bar{\psi}^{(-)}(\vec{x},\vec{k}_{a^{\prime}},s_{a^{\prime}})$ and $\bar{\psi}^{(-)}(\vec{x},\vec{k}_{b'},s_{b'})$ respectively. The four-component wave functions are solutions to the free Dirac equation. The Kronecker product is denoted by $\otimes$, the nucleon-nucleon ($NN$) interaction by $\hat{F}$, and the relativistic bound state wave function by $\phi_{lm_{j}}(\vec{x}\,)$. The Relativistic Impulse Approximation (RIA), which assumes that the $NN$ scattering matrix inside the nuclear medium is identical to that of free $NN$ scattering, is also employed.

    Within the RPWIA framework, the incident and the outgoing protons are treated as free Dirac plane waves given by Equation~\ref{EQ:Free Dirac Plane Wave}, neglecting both the strong nuclear distortion and Coulomb interactions between the protons and the target nucleus \cite{Mello,S.Wyngaardt,PhysRevC.69.024618}. 
    \vspace{-0.55cm}
	\subsection{Relativistic Bound State Wave Functions}\label{Sec:Relativistic boundstate wave functions}
	The bound state proton radial wave functions $\phi_{ljm_{j}}(\vec{x}\,)$ are obtained via self-consistent solutions of the Dirac-Hartree field equations from Quantum Hadrodynamics (QHD) within the Relativistic Mean Field (RMF) approximation \cite{serot1992relativistic}. The wave function is expressed as:\\
	\begin{equation}
		\phi_{ljm_{j}}(\vec{x}\,) =  \frac{1}{x}\sum\limits_{s_{B}} \begin{bmatrix}
			u_{lj}(x)\langle l,m_{j}-s_{B},\frac{1}{2}, s_{B}|j,m_{j}\rangle Y_{l,m_{j}-s_{B}}(\hat{x})\chi_{s_{B}}\\
			iw_{lj}(x)\langle 2j-l,m_{j}-s_{B},\frac{1}{2},s_{B}|j,m_{j}\rangle Y_{2j-l,m_{j}-s_{B}}(\hat{x})\chi_{s_{B}}
		\end{bmatrix},
	\end{equation}\\
	where $u_{lj}$ and $w_{lj}$ represent the upper and lower proton radial wave functions, respectively. The quantum numbers: $l$, $j$, and $m_{j}$ denote the orbital angular momentum, total angular momentum, and its projection of the bound proton. The summation runs over the spin projection $s_{B} = \pm 1/2$ of the bound proton. The brackets $\langle \cdots \rangle$ are the Clebsch-Gordon coefficients, which couple the orbital and spin angular momenta to total angular momentum $j$. The functions $Y_{lm}$ are the spherical harmonics, and $\chi_{s_{B}}$ are two-component Pauli spinors.  \\
    \\
The different RMF models applied to extract the boundstate wave functions are: QHDII, NL3, FSUGold and FSUGold2. QHDII is based on quantum hadrodynamics and includes linear scalar and vector couplings, with its parameters fitted to reproduce bulk properties of nuclear matter \cite{serot1992relativistic}. In contrast, the NL3 model introduces self-interactions of the sigma meson, which improves its ability to describe the ground-state properties of spherical nuclei \cite{PhysRevC.55.540}. FSUGold is an RMF model with additional vector meson self-interactions which improve the description of the symmetry energy and neutron star properties \cite{PhysRevLett.95.122501}. An updated RMF version, FSUGold2, has refined parameters for improved properties of infinite nuclear matter \cite{PhysRevC.90.044305}. Comprehensive model comparisons are beyond the present scope, interested readers should consult the references \cite{serot1992relativistic,PhysRevC.55.540,PhysRevLett.95.122501,PhysRevC.90.044305}.

	\subsection{Nucleon-Nucleon Interaction}\label{Sec:Nucleon-nucleon}
For the $NN$ interaction ($\hat{F}$), we adopt the IA1 representation \cite{Horowitz:1985tw,PhysRevC.37.2032}:
	\begin{equation}
		\hat{F} = \sum_{i=S}^{T}F_i(\lambda^{i{(1)}} \cdot \lambda_{i(2)}),
	\end{equation}
	where the Dirac matrices $\lambda_i$ correspond to scalar, vector, pseudoscalar, axial-vector, and tensor interactions.\\
    \\
We acknowledge the previous studies by Van der Ventel and Hillhouse \cite{PhysRevC.69.024618} where they compared the RPWIA calculations of the ($\vec{p},2\vec{p}$\,) reactions on $^{208}$Pb at 392 MeV, using both IA1 and IA2 parameterizations of the $NN$ interaction, which found that some observables discriminate between the two parameterizations. However, our primary goal is to benchmark RPWIA against the established RDWIA model and not to investigate the dependence on the $NN$ interaction, and to the best of our knowledge, most RDWIA calculations for exclusive knockout reactions (including those in this paper) employ the IA1 representation. To ensure a consistent comparison between RPWIA and RDWIA, we employ IA1 in both frameworks. A systematic study of IA1 versus IA2, within both RPWIA and RDWIA is beyond the scope of this current work, but is planned for future publication. 
\subsection{Observables}\label{Sec:Observables}
The unpolarized triple differential cross section is given by \cite{PhysRevC.68.034608}:
	\begin{equation}
		\sigma = \frac{d^{3}\sigma}{dT_{a^{\prime}}d\Omega_{a^{\prime}}d\Omega_{b}} = S_{lj}\sigma_{\text{calc}},
	\end{equation}
	with
	\begin{equation}
		\sigma_{\text{calc}} = \frac{F_{kin} }{(2s_{a} + 1)(2j + 1 )}\sum\limits_{m_{j},s_{b}}\text{Tr}( TT^{\dagger}),\label{Eq:Cross section}
	\end{equation}
	where $S_{lj}$ represents the spectroscopic factor and $F_{kin}$ is the kinematic factor. The polarization transfer observables, D$_{i'j}$, are defined as:
	\begin{equation}
		\text{D}_{i'j} = \frac{ \text{Tr}(T\sigma_{j}T^{\dagger}\sigma_{i'})}{\text{Tr}(TT^{\dagger})}.\label{Eq: Polarization observables}
	\end{equation}
The transition amplitude in Eq.~\ref{Transition-Amplitude}, is used to construct a $2\times2$ matrix ${T}$ in the spin space of the incident and scattered protons which is used to calculate the cross section and the polarization transfer observables in Eqs.~\ref{Eq:Cross section} and \ref{Eq: Polarization observables} respectively. The $T$ matrix is expressed as:
\begin{equation}
    T = \begin{pmatrix}
        T_{lj}^{s_{a}=\frac{1}{2}, s_{a'}=\frac{1}{2}} & T_{lj}^{s_{a}=-\frac{1}{2}, s_{a'}=+\frac{1}{2}}\\ \\
        T_{lj}^{s_{a}=-\frac{1}{2},s_{a'}=+\frac{1}{2}} & T_{lj}^{s_{a}=-\frac{1}{2},s_{a'}=-\frac{1}{2}}
    \end{pmatrix},\label{T-matrix}
\end{equation}
where $s_{a} = \pm\frac{1}{2}$ and $s_{a'} = \pm\frac{1}{2}$ denote the spin projections of both the incident ($a$) and the outgoing proton ($a'$). The matrix elements $T_{ljm_j}(s_a,s_{a'},s_{b'})$ are related to $T^{s_a,s_{a'}}_{lj}$ through \cite{PhysRevC.68.034608}:
\[
T^{s_a,s_{a'}}_{lj} = \sum_{m_j, s_{b'}} T_{ljm_j}(s_a,s_{a'},s_{b'}).
\]

	%
	\section{Results and Discussion}\label{Sec:Results and Discussion}
This section presents calculations for exclusive proton knockout from the 3s$_{1/2}$ state in $^{208}$Pb at an incident laboratory kinetic energy of 392 MeV. The scattering angles are fixed in coplanar geometry at 32.5$^{\circ}, -50.0^{\circ}$. In Figs.~\ref{fig:Sigma_Ay_results} and \ref{fig:Polarization_transfer_results}, the solid black line (RPWIA) and the dashed black line (RDWIA), both employ the FSUGold2 RMF model for the boundstate wave functions and the gray shaded band represents the spread in the RPWIA predictions from other RMF models (QHDII, NL3, and FSUGold).

	\subsection{Cross Section and Analyzing Power: A Benchmark against Data}
	We begin by comparing the predictions of the RPWIA and RDWIA models with experimental data for the unpolarized triple differential cross section ($\sigma$) and the analyzing power (A$_{y}$) for proton knockout from the 3s$_{1/2}$ state in $^{208}$Pb from Ref.~\cite{Noro2020ExperimentalSO}.
\begin{figure}[htbp]
    \centering
    \includegraphics[width=\textwidth]{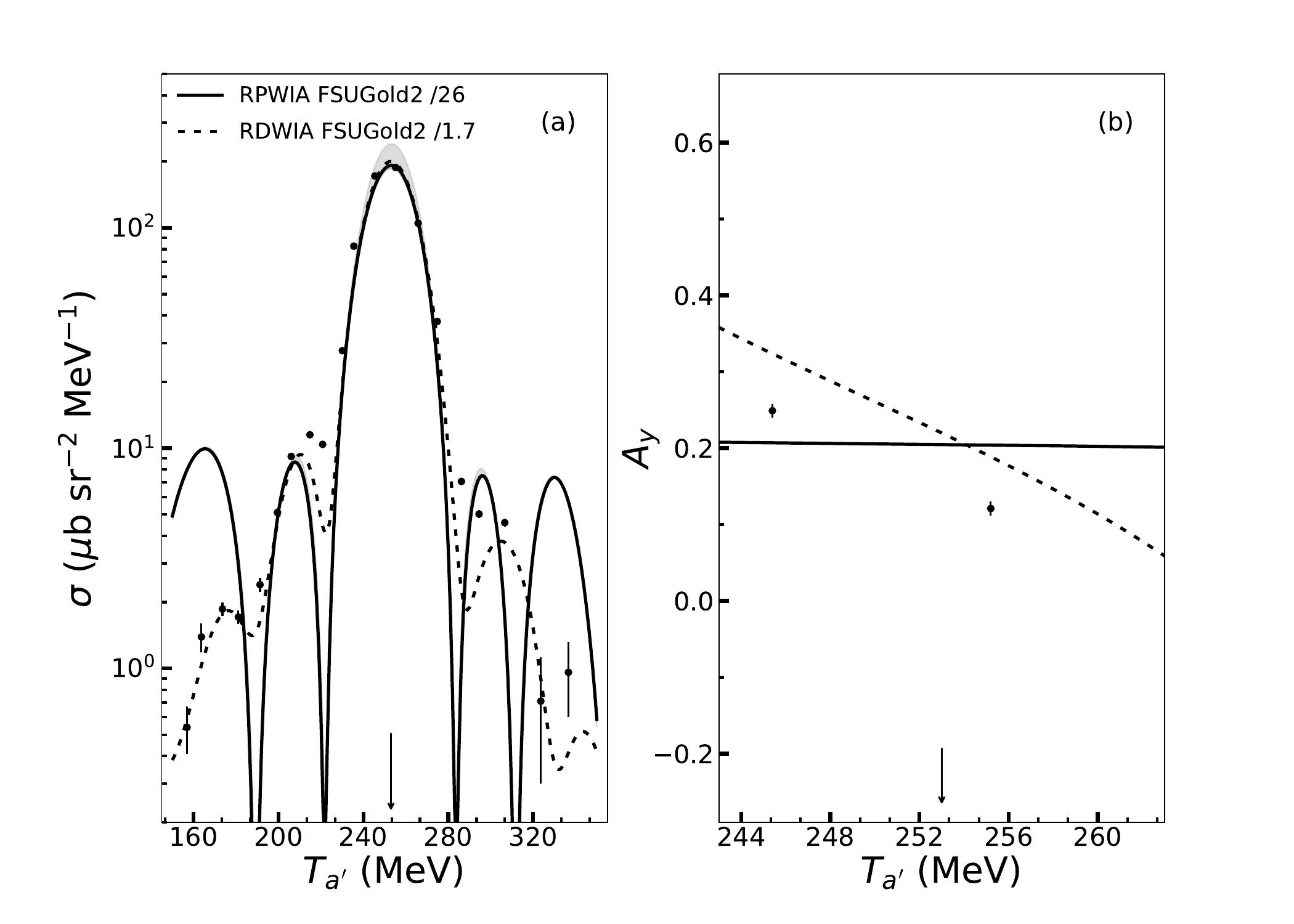}
 \caption{(a) Triple differential cross section $\sigma$, (b) analyzing power $A_{y}$ for the $^{208}$Pb($\vec{p},2\vec{p}$\,) reaction on the 3s$_{1/2}$ state at an incident energy of 392~MeV, and fixed coplanar scattering angles, $32.5^{\circ}, -50.0^{\circ}$, shown as a function of the laboratory kinetic energy $T_{a'}$ of the outgoing proton. The solid (dashed) black line represents the RPWIA (RDWIA) results using the FSUGold2 RMF model. The gray shaded band shows the collective variations in the RPWIA predictions across different RMF models. Experimental data are from Ref.~\cite{Noro2020ExperimentalSO} and the arrows represent the minimum recoil position.}
    \label{fig:Sigma_Ay_results}
\end{figure}
Figure~\ref{fig:Sigma_Ay_results}a shows the energy-sharing distribution of $\sigma$. The RDWIA calculation provides a good qualitative description of the data's shape, requiring only a minor normalization value for a quantitative fit, $S = \sfrac{1}{1.7}$. In stark contrast, the RPWIA model, while reproducing the symmetric shape, dramatically overestimates the magnitude of the cross-section. To match the data, the RPWIA result must be scaled down by a factor of $\sfrac{1}{26}$. This factor is a measure of the total suppression caused by nuclear distortion and absorption—effects that are completely absent within the RPWIA treatment. This result underscores the fundamental inadequacy of the RPWIA for predicting absolute cross-sections.\\
	\\
    The dramatic difference between the spectroscopic factor extracted from the RDWIA ($S = \sfrac{1}{1.7}$) and the normalization factor required for the RPWIA ($\sfrac{1}{26}$) provides a quantitative measure of the importance of distortion effects. While the RDWIA achieves agreement with experimental data through a physically meaningful spectroscopic factor of $\sfrac{1}{1.7}$, the RPWIA requires a phenomenological factor of $\sfrac{1}{26}$ that has no direct physical interpretation. This difference underscores that RPWIA cannot be used for extracting nuclear structure information such as spectroscopic factors, even though it may approximate certain polarization transfer observables within specific kinematic regions.
	Figure \ref{fig:Sigma_Ay_results}b shows the analyzing power A$_{y}$ within the window of $\pm$10 MeV around the peak of the cross section corresponding to zero recoil momentum, the full energy range is shown in Supplemental Material \cite{supp} (Fig.~1).~Here, the performance of the RPWIA model is markedly different.~Within a limited kinematic range (corresponding to low recoil momentum), the RPWIA calculation quantitatively describes the experimental data and agrees closely with the RDWIA prediction. This finding indicates that, while the overall reaction probability is severely suppressed, the spin-dependent aspects of the interaction captured by A$_{y}$ can be modeled reasonably well through the RPWIA approach in this specific kinematic window.

\subsection{Polarization Transfer Observables: A Benchmark against RDWIA}
\begin{figure}[htbp]
				\centering
				\includegraphics[width=1.10\textwidth]{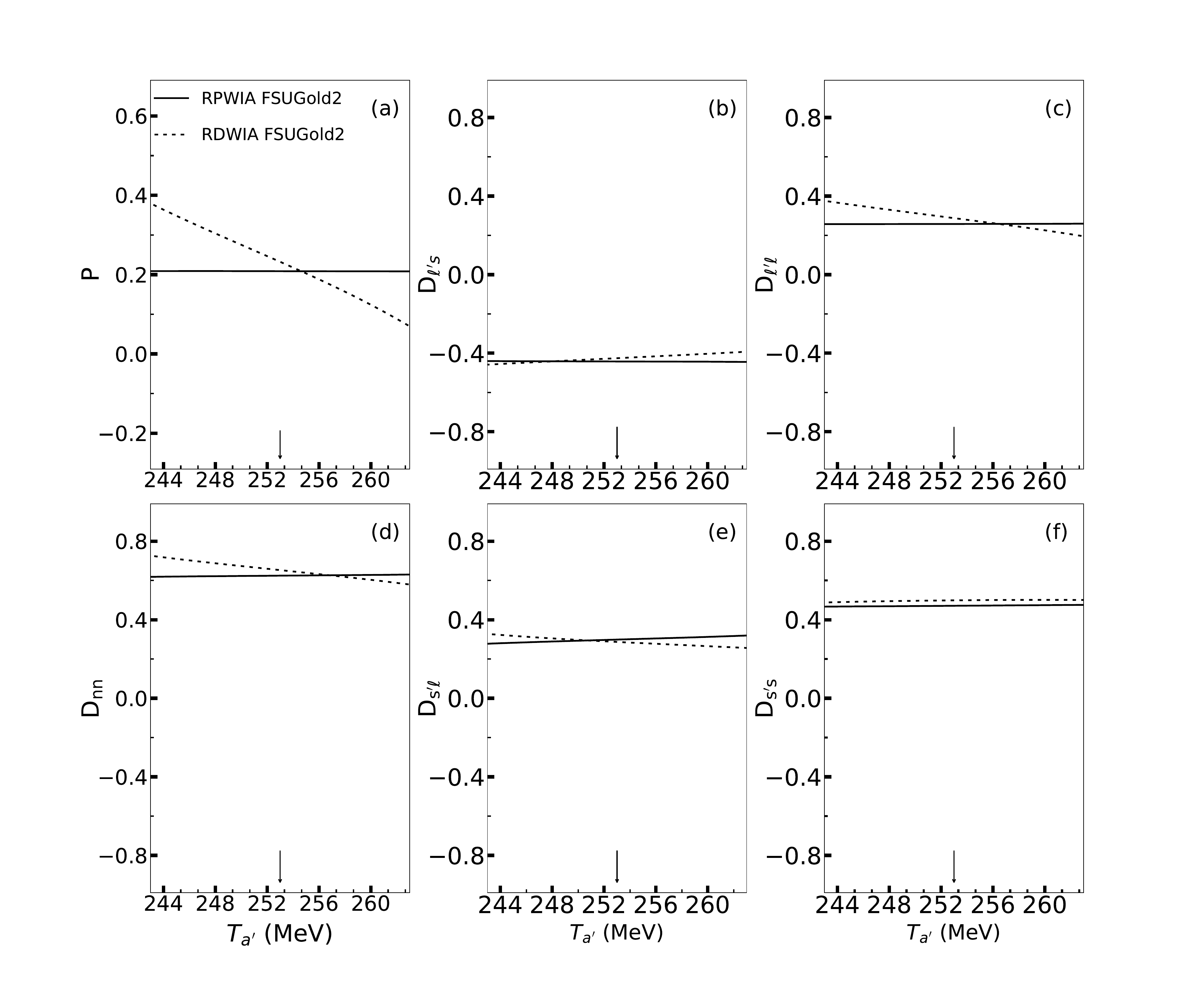}
				 \caption{The complete set of polarization transfer observables (D$_{i'j}$) of the 3s$_{1/2}$ state in $^{208}$Pb with similar RMF model representation as Figure~\ref{fig:Sigma_Ay_results}.}
				\label{fig:Polarization_transfer_results}
			\end{figure}

	As no experimental data currently exist for the complete set of polarization transfer observables (D$_{i'j}$), we benchmark the RPWIA predictions against the established RDWIA model. The results are shown in Figures~\ref{fig:Polarization_transfer_results}a-f.\\
    \\
	Strikingly, within the same $\pm$10 MeV window where A$_{y}$ is well-described, the RPWIA predictions for all D$_{i'j}$ observables are in close agreement with the RDWIA results. This suggests that for these specific spin-transfer observables, the net effect of distortions on the spin dynamics is minimal near the recoilless condition. At these conditions, the momentum transfer to the residual nucleus is essentially zero, since the outgoing protons carry away nearly all the transferred momentum. The optical potentials, which depend on the local momentum of the nucleons, have an insignificant influence on the spin dynamics, since the phase shifts experienced by the incoming and the outgoing waves cancel in spin-dependent ratios. Away from the recoilless regime, however, the residual nucleus acquires non-zero momentum, and the distortion effects become significant, hence leading to noticeable deviations between RPWIA and RDWIA calculations. This behaviour therefore highlights a useful, although limited, application of RPWIA: it can serve as a computationally inexpensive tool for generating initial predictions for polarization transfer observables in this kinematic regime.\\
    \\
We have demonstrated the polarization transfer observables in the limited range of the recoilless condition in Figure~\ref{fig:Polarization_transfer_results} (the $\pm$10 MeV window where RPWIA and RDWIA agree around the peak of the cross section). The results for the full energy range (145 - 355 MeV) are shown in the Supplemental Material \cite{supp} (Fig.~2), where RPWIA and RDWIA deviate outside the $\pm$10 MeV from the peak of the cross section. 
	%
	\subsection{Sensitivity to Nuclear Structure Inputs}
	We investigate the sensitivity of our results to the choice of Relativistic Mean Field (RMF) models (QHDII, NL3, FSUGold, FSUGold2) used to generate the bound-state wave functions. The unpolarized cross section $\sigma$ shown in Figure~\ref{fig:Sigma_Ay_results}a, exhibits noticeable sensitivity to the RMF model. At the recoilless energy ($T_{a'}$ = 253 MeV), the scaled RPWIA cross sections range from 185.9 $\mu$b (FSUGold) to 240.6~$\mu$b (QHDII), with FSUGold2 and NL3 falling between them at 192.4~$\mu$b and 186.8~$\mu$b, respectively, giving a spread of approximately 29\% between the highest and the lowest values. For comparison, the experimental peak cross section is $188.0 \pm 1.3$~$\mu$b. The NL3, FSUGold and FSUGold2 models are within 2.4\% of the experimental data, while QHDII overestimates the cross section by 28\%. However, the spin-dependent observables—both A$_{y}$ (Figure~\ref{fig:Sigma_Ay_results}b) and the complete set of D$_{i'j}$ presented in Figures~\ref{fig:Polarization_transfer_results}a-f show remarkable insensitivity to these changes. This indicates that while the overall probability of the reaction depends on the structural details of the bound state, the spin dynamics governing the polarization observables are predominantly dictated by the free $NN$ interaction and are largely decoupled from the specific nuances of the RMF parameterization, at least for this s-state. We note similar behaviour of the polarization transfer observables for other single-particle states in $^{208}$Pb within the same kinematical conditions; a detailed analysis of these cases will be presented in a future publication.  
\subsection{Spectroscopic Factors and Relativistic Effects}\label{sec:spectroscopic}
The spectroscopic factor ($S$) for the 3s$_{1/2}$ state in $^{208}$Pb extracted from our fully relativistic RDWIA analysis is $S = 0.61(14)$, obtained by normalizing the theoretical cross-section to experimental data following procedures in Ref.~\cite{Noro2020ExperimentalSO}. This value, is presented alongside comparative results in Table~\ref{tab:spect_factors}.\\
\\
The NRDWIA analyses in Ref.~\cite{Noro2020ExperimentalSO}, yielded larger values ($S$ $\approx$ 1.6 - 1.8) for the same reaction on $^{208}$Pb. As noted in Ref.~\cite{Noro2020ExperimentalSO}, the tendency of the NRDWIA to overestimate spectroscopic factors is not universal since it has provided $S$ factors which are consistent with those extracted from ($e,e'p$) measurements for lighter nuclei, i.e., $^{12}$C, $^{16}$O, and $^{40}$Ca. The discrepancy appears for a heavy nucleus like $^{208}$Pb, where the relativistic and distortion effects play a significant role.
\begin{table}[htbp]
    \centering
    \caption{Spectroscopic factors for the 3s$_{1/2}$ state in $^{208}$Pb from various reaction mechanisms. The present RDWIA analysis yields $S = 0.61(14)$, which is lower than values obtained from other approaches.}
    \label{tab:spect_factors}
    \begin{tabular}{llcl}
        \hline\hline
        {Reaction } & {Model/Reference} & {Spectroscopic Factor $S$ } & {Notes} \\
        \hline
        $(\vec{p},2\vec{p}\,)$ & Present RDWIA work & 0.61(14) & Fully relativistic treatment \\
        $(p,2p)$ & EDAD1 \cite{Noro2020ExperimentalSO} & 1.64(32) & Semi-relativistic \\
        $(p,2p)$ & Timora \cite{Noro2020ExperimentalSO} & 1.77(41) & Semi-relativistic \\
        $(e,e'p)$ & Various \cite{Noro2020ExperimentalSO,WAKASA201732,KRAMER2001267} & 0.98(09) & Electron scattering \\
        $(d,^{3}\mathrm{He})$ & Transfer reactions \cite{WAKASA201732,KRAMER2001267} & 1.5  & Nuclear transfer \\
        \hline\hline
    \end{tabular}
\end{table}\\
The reduction in spectroscopic strength extracted from the RDWIA framework compared to other analyses ($S = 0.61$ versus $S \approx 1.6-1.8$ from earlier $(\vec{p},2\vec{p}\,)$ studies) may be influenced by several factors inherent to the consistent relativistic treatment:
\begin{enumerate}
    \item \textbf{Strong relativistic potentials:} The RDWIA framework includes both scalar and vector potentials of substantial magnitude ($\approx300 - 400$ MeV in the nuclear interior \cite{PhysRevC.47.297}), which may strongly distort the Dirac spinors of the bound and scattering states.
    \item \textbf{Enhanced wave function suppression:} These strong potentials may lead to additional suppression of the overlap between initial and final states beyond what is captured in non-relativistic or semi-relativistic frameworks.
    \item \textbf{Consistency with relativistic dynamics:} The lower $S$ value may reflect a more complete treatment of relativistic dynamics throughout the reaction process.
\end{enumerate}
It is important to exercise caution when interpreting the differences in the $S$ factors from different reaction models. The RDWIA model itself faces several challenges, including the proper treatment of the recoil effects, sensitivity to the choice of relativistic optical potentials, and limited validation for lighter nuclei where NRDWIA has been extensively tested. Consequently, while our RDWIA result for $^{208}$Pb is physically plausible, a more comprehensive assessment for a wider range of nuclei and kinematics is required before making conclusions about the relative performance of relativistic (RDWIA) and non-relativistic (NRDWIA) frameworks. Even though the results of a fully relativistic treatment may yield lower $S$ factor for the 3s$_{1/2}$ state of $^{208}$Pb, this should not be interpreted as a claim of RDWIA's superiority over NRDWIA.\\
\\
Our discussion reinforces the message of the paper regarding RPWIA's limitations. The large difference between the factor required to scale the RPWIA results to data ($\sfrac{1}{26}$) and that of the RDWIA (0.61) demonstrates that distortion effects are important for quantitative descriptions of the absolute cross sections. Even for sophisticated distorted-wave models (RDWIA and NRDWIA), the extracted $S$ factors vary and dependent on the type of theoretical framework used. This highlights the need to establish the use of validated reaction models when extracting nuclear structure information from knockout reactions. \\
\\
The discrepancy between our RDWIA result and previous $(p,2p)$ analyses highlights the significant impact of theoretical framework choices on extracted spectroscopic factors. Notably, earlier $(p,2p)$ analyses employed hybrid approaches (e.g., non-relativistic distorted waves with relativistic bound states) \cite{Noro2020ExperimentalSO}, which may not fully account for the strong relativistic distortion effects present in heavy nuclei like $^{208}$Pb.\\
\\
The factor of $\approx16$ difference between the RDWIA normalization ($S = 0.61$) and the RPWIA normalization ($\sfrac{1}{26}$) is smaller than the $\approx43$ difference when comparing to NRDWIA analyses. This emphasizes that:
\begin{itemize}
    \item Even sophisticated models like RDWIA show significant variation in extracted spectroscopic factors depending on implementation details and relativistic consistency.
    \item RPWIA's failure is extreme when compared to either relativistic or non-relativistic treatments, requiring a normalization factor $\sfrac{1}{26}$ that has no relation to physically meaningful spectroscopic strengths.
    \item The separation between what RPWIA can do (approximate certain polarization observables in specific kinematics) and what it cannot do (extract nuclear structure information) is now quantitatively established through this comparison.
\end{itemize}
\newpage
The systematically lower spectroscopic factor extracted from the RDWIA framework ($S=0.61$) compared to earlier semi-relativistic $(\vec{p},2\vec{p}\,)$ analyses ($S\approx1.6-1.8$) demonstrates the critical importance of maintaining relativistic consistency throughout the reaction model. This consistency appears to yield spectroscopic strengths that align more closely with values from fundamentally different reaction mechanisms like $(e,e'p)$, potentially offering a more unified description of nuclear structure across diverse experimental probes. This finding underscores that the precise value of extracted spectroscopic factors remains strongly dependent on the completeness of the theoretical framework, particularly for heavy nuclei where relativistic effects are significant.
\section{Conclusion}\label{sec:conclusion}
We have conducted a benchmark of the Relativistic Plane Wave Impulse Approximation (RPWIA) against the established Relativistic Distorted Wave Impulse Approximation (RDWIA) for exclusive $(\vec{p},2\vec{p}\,)$ knockout reactions on the 3s$_{1/2}$ state in $^{208}$Pb at 392 MeV. Our key findings are:
\begin{enumerate}
    \item RPWIA fails for absolute cross-sections, requiring an unphysical normalization factor of $\sfrac{1}{26}$ compared to the spectroscopic factor $S=0.61(14)$ extracted from RDWIA.
    \item Within $\pm10$ MeV of the recoilless condition, RPWIA quantitatively describes both the analyzing power ($A_y$) data and RDWIA predictions for all polarization transfer observables D$_{i'j}$.
    \item Polarization observables in this kinematic window are insensitive to the choice of RMF model, governed primarily by the free $NN$ interaction.
    \item The spectroscopic factor from RDWIA ($S = 0.61$) is lower than values yielded by the NRDWIA analyses ($S \approx 1.6 - 1.8$). This highlights the sensitivity of the extracted $S$ factors to the reaction model for heavy nuclei.
    
\end{enumerate}
Therefore, RPWIA has limited but specific utility: it provides computationally simple, accurate estimates for polarization observables near zero recoil, making it suitable for exploratory studies with radioactive beams where optical potentials are unknown. For spectroscopic factor extraction or full kinematic analyses, RDWIA and NRDWIA remain essential. This work establishes clear boundaries for applying simplified models in polarized knockout reactions.
\section*{Acknowledgement}
\noindent
TM acknowledges financial support from the Botswana International University of Science and Technology (BIUST) postgraduate funding, under grant number S00526. The authors thank Prof. J. Piekarewicz for providing the code for extracting the RMF properties of infinite nuclear matter. 

	\bibliographystyle{apsrev4-2}
	\bibliography{references}
	
\end{document}